\documentclass[journal=nalefd,manuscript=article]{achemso}
\usepackage{chemformula} 
\usepackage[T1]{fontenc} 
\usepackage[version=3]{mhchem} 
\usepackage{bm}
\usepackage{siunitx} 
\usepackage[normalem]{ulem}
\usepackage{color}
\usepackage{xcolor} 

\author{Ralph L. Stoop}
\affiliation[Universitat de Barcelona]{Departament de F\'isica de la Mat\`eria Condensada, Universitat de Barcelona, Av. Diagonal 647, 08028 Barcelona, Spain}
\author{Arthur V. Straube}
\affiliation[Freie Universit\"at]{Freie Universit\"at Berlin, Department of Mathematics and Computer Science, Germany}
\alsoaffiliation[Universitat de Barcelona]{Departament de F\'isica de la Mat\`eria Condensada, Universitat de Barcelona, Av. Diagonal 647, 08028 Barcelona, Spain}
\author{Pietro Tierno}
\affiliation[Universitat de Barcelona]{Departament de F\'isica de la Mat\`eria Condensada, Universitat de Barcelona, Av. Diagonal 647, 08028 Barcelona, Spain}
\alsoaffiliation[IN2UB]{Institut de Nanoci\`encia i Nanotecnologia,
Universitat de Barcelona, Barcelona, Spain}
\alsoaffiliation[UBICS]{Universitat de Barcelona Institute of Complex Systems (UBICS), Barcelona, Spain}
\email{ptierno@ub.edu}
\title{Enhancing nanoparticle diffusion on a unidirectional domain wall magnetic ratchet}
\begin{document}


\begin{abstract}
The performance of 
nanoscale magnetic devices
is often limited by the presence of 
thermal fluctuations,
while in micro-nanofluidic applications
the same fluctuations may be used to 
spread reactants or drugs. 
Here we demonstrate the controlled motion and the 
enhancement of diffusion of magnetic nanoparticles 
that are manipulated and driven 
across a series of Bloch walls within an
epitaxially grown ferrite garnet film.
We use a rotating magnetic field to generate a traveling wave potential that unidirectionally transports the nanoparticles at a frequency tunable speed. 
Strikingly, we find an enhancement of diffusion 
along the propulsion direction and a frequency dependent diffusion coefficient that can be precisely controlled by varying the system parameters.  
To explain the reported phenomena, 
we develop a theoretical approach that 
shows a fair agreement with the experimental data 
enabling an exact analytical expression
for the enhanced diffusivity 
above the magnetically modulated periodic landscape.
Our technique 
to control thermal fluctuations of 
driven magnetic nanoparticles represents a 
versatile and powerful way to programmably 
transport magnetic colloidal matter in a fluid,
opening the doors to different 
fluidic applications based on 
exploiting magnetic 
domain wall ratchets.
\end{abstract}
The ratchet effect emerged in the past as a powerful way 
to transport matter at the micro and nanoscale, taking 
advantage of Brownian motion~\cite{Han09,Rei02}.
The success of such concept comes from the possibility of using thermal fluctuations to obtain useful work out of a thermodynamic system,
although such fluctuations  
produce noise, heat and randomize the motion of nanoparticles which limits 
the efficiency of any device operating at such scale~\cite{Eri13,Crid06,San17}. 
Reducing or controlling thermal fluctuations in 
nanoparticle systems  
may present different technological advantages 
apart from providing important fundamental insight 
into the dynamics and interactions of matter at such scale. 
In the first case, diffusion can be used as a mean for mixing streams of fluids or for spreading reactants, drugs and biological species~\cite{Wal11,Skau15} in micro- and nano-fluidic applications~\cite{Squ05,Sch08,Val2012}.
On a more fundamental level, the search for 
strategies that enable controlling diffusion and 
noise has fascinated scientists
for long time, 
since the pioneering seminar of Richard Feynman
on Brownian ratchet~\cite{Fey63}. 

In a typical ratchet system 
the random fluctuations of 
nanoparticles can be rectified into a directed motion by an external potential.
In overdamped systems, different ratchet mechanisms can be sorted in two general classes, depending on the nature of the external potential~\cite{Rei02}. One class refers to the tilting ratchets that are typically specified by a stationary, spatially asymmetric potential landscape often accompanied by an external force. The asymmetry of the landscape is able to rectify thermal fluctuations into net motion, and the force, which determines the tilt of the total potential, can be used to further enhance the particle current. Such a ratchet scheme can be realized by using, for example, the walls or barriers in a microfluidic device~\cite{Mat03}. Another broad class are the pulsating ratchets that are characterized by a time-dependent landscape, in which the net particle flux arises from the periodic or random evolution of the landscape. Their particular subclasses are the flashing ratchet with the landscape switching between two states, and the traveling wave ratchet, where the landscape translates at a given speed. The latter allows dragging the particles that are trapped in the energy minima of the moving landscape.

The generation of translating potential landscapes usually requires an external field, which makes it appealing due to its programmable nature, that enables to remotely control, direct and reverse the flux of particles at will~\cite{Rei17}.  
Recent realizations 
in this direction include the use of electric~\cite{Jul94,Mar02,Kim18,Wei17,Ska18}, optical~\cite{Fau95,Lee05,Sha2010,Arz17} or magnetic~\cite{Gun05,Yel05,Tie2007} fields to transport microscopic particles.  
However, most of the proposed approaches 
remain difficult to apply at the nanoscale,
due to the large field gradients required 
to overcome thermal forces.
Plasmonic landscapes have been proposed 
as a successful strategy~\cite{Rig08,Sha2010,Kua13}, however the
created patterns are often composed of 
fixed lithographic gold structures that 
cannot be easily changed by an external control. 
An alternative solution is the use of patterned magnetic substrates 
that present high contrast in the magnetic susceptibility on the particle scale and thus generate strong and localized field gradients~\cite{Hlh03,Ehr11,Don10,Rap12}.  
While the trapping of magnetic nanoscale matter was recently demonstrated with lithographic~\cite{Dem2013} and epithaxial~\cite{Tie2016} films, 
the complete control of particle transport, speed and diffusion  
within the same functional platform 
still remains a challenging issue.  
  
In this article we demonstrate all these features by using a periodic 
patterned substrate, i.e. a 
uniaxial ferrite garnet
film (FGF), composed of a series of 
mobile magnetic Bloch walls. 
Controlling the dynamics of magnetic domain walls
in thin films has recently led to 
novel applications in disparate fields, including spintronics,~\cite{Sev2014} 
logic devices~\cite{Fra2012,Phu2015},
nanowires~\cite{Gu2007,Xin2011}
and ultracold atoms.~\cite{Wes2012}
Here we use such magnetic walls to achieve
the controlled transport and, in particular, the diffusion enhancement of magnetic nanoparticles. 
The application of an external rotating magnetic field induces the unidirectional translation of the otherwise stationary magnetic landscape generated by the FGF. Due to the magnetic interaction with the nanoparticles, the translation of the landscape is converted into directed particle motion. This technique allows to rapidly traps and steers magnetic nanoparticles deposited on top of it thus, with in situ real time control. Moreover, our magnetic ratchet allows us to precisely control also the diffusivity of the nanoparticles, to enhance, to suppress it, by simply changing the driving frequency of the applied field.

\begin{figure}[h]
\begin{center}
\includegraphics[width=\columnwidth,keepaspectratio]{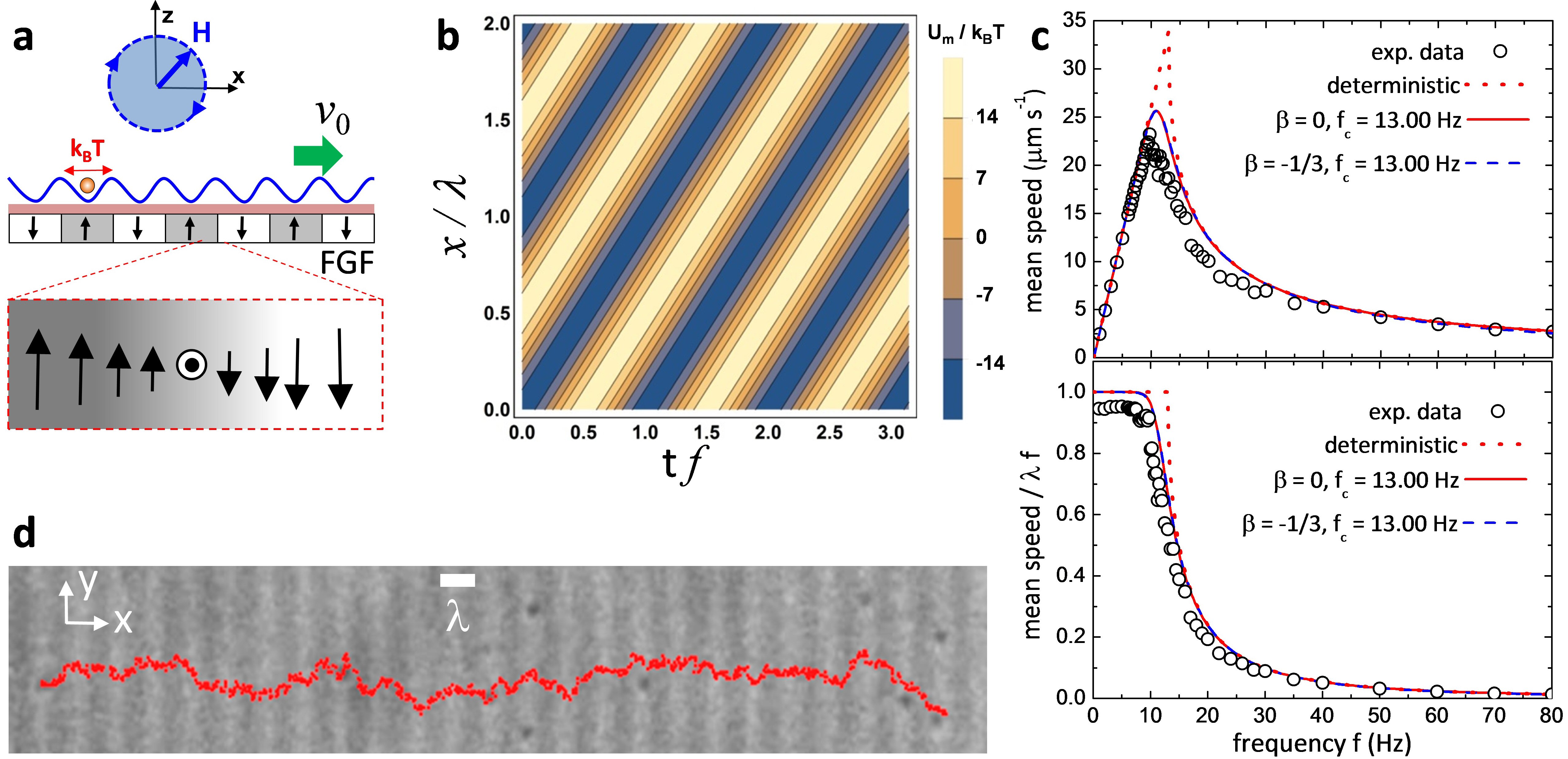}
\caption{(a) Schematic illustrating a single magnetic nanoparticle 
above an FGF subjected to a rotating magnetic field polarized in the ($x,z$) plane. The applied field helps to generate a periodic potential (blue line) that translates at constant speed $v_0=\lambda f$
and transports the nanoparticle. Bottom illustrates 
the rotation of magnetization in a Bloch wall.
(b) Color coded energy landscape  ($U_m/k_BT$) versus particle position
and time calculated for a $\SI{270}{\nano\meter}$ nanoparticle (effective magnetic volume susceptibility $\chi=1$, elevation from surface $z= \SI{1.34}{\micro\meter}$, $f= \SI{2}{\hertz}$), low energies in blue and high energies in yellow. The particle transport occurs consistently with the blue regions.  (c)  
Particle mean speed $\langle v \rangle$ across the BWs versus frequency $f$ (top) and normalized with respect to $\lambda f$ (bottom), for a  $\SI{270}{\nano\meter}$ nanoparticle subjected to a rotating field with amplitude $H_0 = \SI{1200}{\ampere\per\meter}$. The experimental data (empty circles $\beta = -1/3$) are
plotted together with the theoretical lines: The dashed line
shows the deterministic limit, Equation \ref{vav-det}, while the red solid line also accounts for the effect of thermal fluctuations, Equation \ref{vav}, see the Materials and Methods section.
(d) Polarization microscope image showing a $\SI{270}{\nano\meter}$ particle trajectory transported by the rotating field at an average speed $\langle v \rangle = \SI{5.2}{\micro\meter\per\second}$. The field parameters are  $f=\SI{2}{\hertz}$, $H_0 = \SI{1200}{\ampere\per\meter}$.
The trajectory (red line) is superimposed to the image, $\lambda = \SI{2.6}{\micro\meter}$, see VideoS1 in the Supporting Information.}
\label{fig1}
\end{center}
\end{figure}

Our magnetic ratchet scheme is shown in Figure~\ref{fig1}(a) and it is based on a series of Bloch walls (BWs),
narrow ($\sim \SI{20}{\nano\meter}$ wide) transition regions where
the film magnetization
changes orientation by performing a 
180 degrees rotation in the particle plane $(x,y)$.
Such BWs emerge spontaneously in a 
single crystal FGF film after an epitaxial growth process and they generate strong local stray fields on the surface.
In this study we use an FGF of composition 
Y$_{2.5}$Bi$_{0.5}$Fe$_{5-q}$Ga$_{q}$O$_{12}$
($q=0.5-1$),
that was grown by dipping liquid phase epitaxy
on a $\langle 111 \rangle$ oriented single crystal
gadolinium  gallium  garnet (Gd$_3$Ga$_5$O$_{12}$)
substrate. In absence of external field,  
the BWs in the FGF are equally spaced at a distance $\lambda/2$ 
being $\lambda = \SI{2.6}{\micro\meter}$ the spatial periodicity,
and separates domains 
of opposite magnetization, with saturation magnetization $M_\text{s} = \SI{1.3e4}{\ampere\per\meter}$. 

Above this film we demonstrate the controlled transport of three 
three types of paramagnetic polystyrene nanoparticles with
diameters
$d=$ \SIlist{540;360;270}{\nano\meter},
characterized by a $\sim 40 \% \, {\rm wt.}$
iron oxide content (Microparticles GmbH).
When placed above the FGF, the nanoparticles 
show simple diffusive dynamics and are able to easily pass the BWs due to the presence of a $\SI{1}{\micro\meter}$ thick polymer coating that prevents sticking to the FGF surface, see the Materials and Methods section. 
We apply an external magnetic field rotating in the $(x,z)$ plane with frequency $f$ and 
amplitudes $(H_x,H_z)=H_0(\sqrt{1+\beta}, \sqrt{1-\beta})$, such that $\bm{H}^\text{ac}(t)= (H_x \cos {(2\pi f t)}, 0, -H_z \sin {(2\pi f t)} )$, Figure~\ref{fig1}(a).
The polarization of the applied field is slightly elliptical, $H_x\ne H_z$, as characterized by the amplitude $H_0=\sqrt{(H_x^2 +H_z^2)/2}$ and an ellipticity parameter $\beta = (H_x^2-H_z^2)/(H_x^2+H_z^2)$. 
The value of $\beta=-1/3$ is chosen to minimize dipolar interactions between nearest particles~\cite{Str14}, as they can promote the undesired chaining
at high density~\cite{Tierno2016}. Although for self-consistency we account for this small ellipticity in a full numerical model, a more tractable semi-analytical model for the special case of $\beta=0$ is shown to work quantitatively well. 

The applied field modulates the stray magnetic field generated by the FGF, leading to a spatially periodic magnetic energy landscape $U_\text{m}(x,t) = -U_0\cos(k(x-v_0t))$ that translates at a constant speed, $v_0 = \lambda f$, Figure~\ref{fig1}(a), see Materials and Methods section for further details and definition of $U_0$ and $k$. Its evolution at the elevation corresponding to the \SI{270}{\nano\meter} particles is shown in Figure~\ref{fig1}(b).
One may intuitively expect that the particles would follow the energy minima whose locations are given by the blue regions, and travel the distance of one wavelength during one time period. However, the nature of motion and the velocity of mean drift across the BWs, 
\begin{align}
\langle v \rangle=\lim_{t\to\infty}\frac{\langle X(t)\rangle}{t},
\quad X(t)=x(t)-x(0),
\label{DEF:v-mean}
\end{align}
depend on the external frequency, as becomes evident for the simple case of no thermal fluctuations, see Equation~\ref{vav-det}.
The variation of $\langle v \rangle$ with the  
frequency $f$ is also shown in Figure~\ref{fig1}(c), 
where the experimental data (open symbols) 
are plotted along with the theoretical predictions (dashed and solid lines) for a \SI{270}{\nano\meter} particle.  

As it follows from our model (see Equation~\ref{vav-det} in Materials and Methods), and as confirmed by the experimental data, there exist two different dynamic states separated by the critical frequency $f_\text{c}$. At low frequencies, here $f< f_\text{c} \approx \SI{13}{\hertz}$, the nanoparticles indeed translate consistently with the landscape with the maximum possible speed, $\dot x(t)=\langle v \rangle = v_0=\lambda f$; see  the range of perfectly linear increase of $\langle v \rangle$ with $f$ in Figure~\ref{fig1}(c). We also note that at any $0 < f < f_\text{c}$ in this ``locked'' regime, the particle lags behind the minimum of the potential energy landscape $U_\text{m}$, whose relative position is determined by minimizing the potential given by Equation \ref{potV}. For $f>f_\text{c}$, the particle is unable to move together with the landscape. It starts to slip and cannot remain localized within the same minimum. In this ``sliding'' regime, every time period the particle covers a distance smaller than $\lambda$, resulting in a reduction of the mean speed. At high frequencies $f$, the speed of mean drift decays as $\langle v \rangle \simeq \lambda f_\text{c}^2/(2f)$. 
Note that at low and high frequencies, thus far away from $f_\text{c}$, the predictions of the model with thermal fluctuations and the experimental data are highly consistent with the deterministic (zero temperature) results given by Equation (4). However, the deterministic model overshot the experimental data close to $f_\text{c}$. Therefore, we conclude that thermal fluctuations start to significantly affect the particle speed close to the critical frequency, blurring the sharp transition from the locked to the sliding dynamics and effectively decreasing the value of $f_\text{c}$ as compared to the relative to the deterministic model.

\begin{figure}[ht]
\begin{center}
\includegraphics[width=0.8\columnwidth,keepaspectratio]{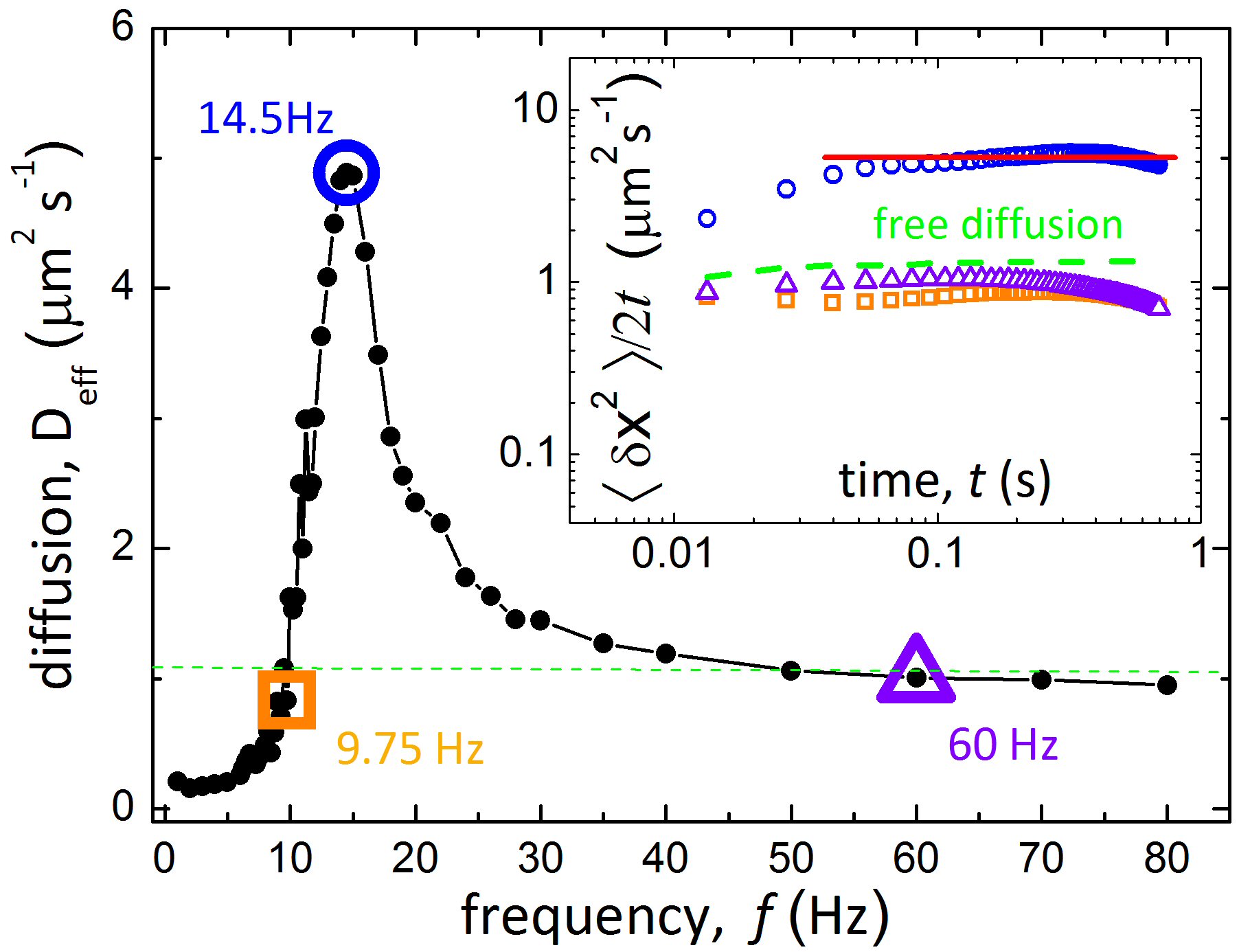}
\caption{Effective diffusion coefficient $D_\text{eff}$ 
measured along the propulsion direction versus driving frequency $f$ for particles with $d=\SI{270}{\nano\meter}$. 
For all data, the particles were driven above the ratchet by a rotating field 
with amplitude $H_0=\SI{1200}{\ampere\per\meter}$ and $\beta=-1/3$.
The green dashed line corresponds to the 
free diffusion coefficient measured 
above the FGF and in absence of applied field. 
The small inset displays the mean square displacements (MSDs)
divided by the time, $\langle \delta x^2 \rangle/(2t)$
for three different frequencies with locations 
indicated by open symbols in the main plot.
The continuous red line through the $f=14.5\rm{Hz}$ data shows a linear fit to calculate $D_{eff}$.}
\label{fig2}
\end{center}
\end{figure}

While our domain wall magnetic ratchet enables full control over the mean speed of particle moving across BWs, the instant position of the particles is 
affected by unavoidable thermal fluctuations, as evidenced by the particle trajectory in Figure~\ref{fig1}(d).
The role of such fluctuations is typically quantified by measuring the mean squared displacement (MSD), which can be calculated from the positions of the nanoparticles. 
Since we are interested in the general 
effect of how the diffusive properties of a nanoparticle 
are influenced by the ratchet mechanism, 
we focus on investigating these quantities 
along the propulsion direction, namely the $x$ axis. Because of the non vanishing mean drift, we define the MSD as the variance of the corresponding particle position, $\sigma_x^2(t)=\langle \delta x^2(t) \rangle \sim t^{\alpha}$, where $\delta x(t) = X(t) - \langle X(t)\rangle$.
For our statistical analysis we averaged over more than $\sim 50$ independent experimentally measured trajectories, for each of which we subtracted its  mean drift, $\langle X(t)\rangle = \langle v\rangle t$.
The exponent of the power law $\alpha$ can be used to distinguish the diffusive $\alpha = 1$ from anomalous (sub-diffusive $\alpha <1$ or super-diffusive $\alpha >1$) dynamics. Our case corresponds to normal diffusion, with an effective diffusion coefficient across the BWs evaluated as
\begin{align}
D_\text{eff}=\lim_{t \to \infty} \frac{\sigma^2_x(t)}{2t}.
\label{DEF:Deff}
\end{align}
In Figure~\ref{fig2} we show that effectively, the magnetic ratchet provides a strong enhancement of the diffusion coefficient at an optimal frequency. At low frequencies, the diffusive motion of the nanoparticles with $d=\SI{270}{\nano\meter}$ corresponds to a $D_\text{eff}$ much lower than the diffusivity in absence of the external field, $D_0=\SI{1.0}{\micro\meter\squared\per\second}$. At high frequencies, the diffusivity becomes close to $D_0$. At intermediate frequencies, the effective diffusivity increases (note that $D_\text{eff}\approx 0.6 D_0$ for $f=\SI{11.5}{\hertz}$ and $f=\SI{70}{\hertz}$) with an enhancement of almost one order of magnitude relative to free diffusivity, $D_\text{eff} \approx 8.5 D_0$, at a frequency of $f=\SI{15.5}{\hertz}$. 
In particular, as shown in the inset of Figure 2, after subtracting the mean drift we calculate the effective diffusion coefficient using Eq. 2 and performing a linear fit only for the region of data where the mean square displacement displays a power law behavior with $\alpha = 1$. This regime is always found in the long-time limit. Overall, the dependence of $D_\text{eff}$ on $f$
follows a well defined trend, characterized by an initial sharp 
raise above a value of $f \approx \SI{8}{\hertz}$, and an exponential-like decay above a peak at $\SI{15.5}{\hertz}$, which is close to the critical frequency value measured for this size of nanoparticle, $f_\text{c}= \SI{13.4}{\hertz}$.

\begin{figure}[ht]
\begin{center}
\includegraphics[width=\columnwidth,keepaspectratio]{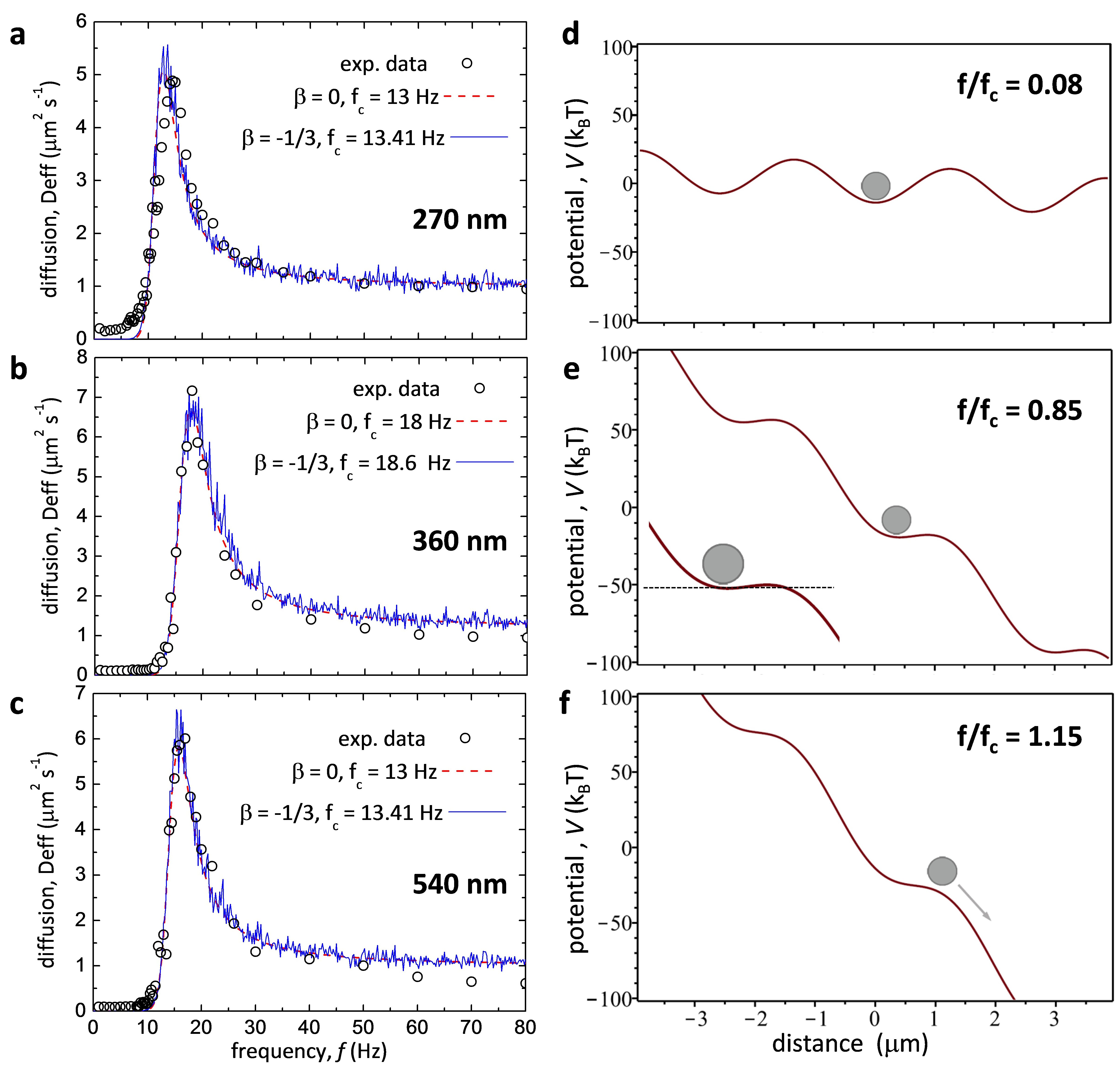}
\caption{(a-c) Effective diffusion coefficient $D_\text{eff}$ 
versus frequency $f$ for three types of particles with diameters $\SI{270}{\nano\meter}$ (a), $\SI{3600}{\nano\meter}$ (b) 
and $\SI{540}{\nano\meter}$ (c). 
All panels show the experimental data (open circles),
predictions of the reduced theoretical model with $\beta=0$ according to Equation \ref{deff} (red line) and estimates of the effective diffusion coefficient from numerical simulations of the full model with $\beta=-1/3$, Equation \ref{LE-gen} (blue lines).
The experimental parameters used are 
$H_0=\SI{1200}{\ampere\per\meter}$ and $\beta = -1/3$
that reflect the values used for the simulations. 
For the three types of particles we also indicate the critical frequencies $f_\text{c}$ used for the model (red) and 
simulation (blue).
(d-f) Magnetic potential $V(u)$ in units of thermal energy $k_B T$ in the moving reference frame (Equation $5$) evaluated for a $270 \rm{nm}$ particle and at a frequency of  $f=1 \rm{Hz}$ (d), $f=11 \rm{Hz}$ (e) 
and $f=15 \rm{Hz}$ (f). Here the critical frequency is $f_c=13 \rm{Hz}$.}
\label{fig3}
\end{center}
\end{figure}

The observed enhancement of diffusion 
can be explained by formulating 
a reduced theoretical model that explicitly takes into account 
the interaction of the nanoparticle with the magnetic landscape 
generated by the FGF surface and the thermal noise.
Details of the derivation are given in the Materials and Methods section. 
In the reference frame moving with the magnetic potential, the behavior of the nanoparticle is equivalent to the motion in an effective tilted periodic potential $V(u)$ as given by Equation \ref{potV}. For such a potential, the effective diffusion coefficient, defined via Equation \ref{DEF:Deff}, can be expressed as Equation \ref{deff}, which admits an accurate numeric evaluation. We used Equation~\ref{deff} in Figures~\ref{fig3}(a)-\ref{fig3}(c) to fit the experimental data for the three types of nanoparticles, and find that our theoretical approach captures very well the observed diffusion 
enhancement for all cases.

The explicit form of the effective potential, Equation \ref{potV}, with the tilt $\propto f$ and the amplitude of the landscape $\propto f_\text{c}$ admits an intuitively clear interpretation of the observed frequency-dependent diffusive regimes, as also illustrated in Figures 3(d)-3(f). Indeed, at subcritical frequencies, $f \le f_\text{c}$, the potential barrier the particle needs to overcome to escape from a minimum drops with the frequency as $\Delta V /k_\text{B}T \approx \Delta \upsilon_\text{m}(1-f/f_\text{c})^{3/2}$, where $\Delta \upsilon_\text{m}=\lambda^2f_\text{c}/(\pi D_0)$ is double the amplitude of the landscape relative to thermal energy, cf. Equation~\ref{potV}.
At low frequencies ($f \ll f_\text{c}$), the tilt is negligible and the particle is strongly trapped in a minimum of the magnetic landscape, as shown in Figure 3(d). Since in this case the potential barrier is maximum, $\Delta V /k_\text{B}T \approx \Delta \upsilon_\text{m}$, and is too high for the particle to escape, the effective diffusivity is nearly vanishing, $D_\text{eff} \ll D_0$. At intermediate frequencies, as $f$ tends to $f_\text{c}$ but remains smaller than $f_\text{c}$, the potential barrier decreases with $f$ and disappears at $f=f_\text{c}$. In this frequency range, the barriers become progressively more accessible for the Brownian particle, 
explaining the observed significant enhancement of the effective diffusivity at frequencies $f \approx f_\text{c}$, see Figure 3(e). Beyond $f=f_\text{c}$, there exits no minima in the potential, and at high frequencies $f \gg f_\text{c}$, the particle does not feel the landscape,
as shown in Figure 3(f).
As a result, the diffusive motion occurs effectively at a constant force $\propto f$ and therefore corresponds to free diffusion, when ideally $D_\text{eff}=D_0$. 

Further, since the reduced model was developed for the zero field ellipticity, $\beta=0$, we have performed control Brownian dynamic simulations of the 
full system with $\beta=-1/3$ to confirm the negligible effect 
of $\beta$ on the basic physics. The simulation (blue lines) agree very well with the model and the experimental data. The only difference is that the system with $\beta=-1/3$ is characterized by slightly higher values of the critical frequency compared to the case $\beta=0$, which is in agreement with our earlier observation that the value of $f_\text{c}$ at any $\beta \ne 0$ is generally smaller than that for $\beta=0$ at otherwise identical conditions \cite{Str14}.
Taken together, our findings show that each type of 
nanoparticle investigated exhibit an enhanced diffusive behavior, with the highest diffusion coefficient measured for the intermediate size, $d=\SI{360}{\nano\meter}$. 
This outcome can be understood by considering the balance between two opposite effects. First, reducing the nanoparticle diameter $d$ increases thermal fluctuations and its free diffusion coefficient $D_0=k_B T/\zeta$, where $\zeta$ is the friction coefficient of a spherical particle immersed in water. Note that if in the bulk $\zeta=3 \pi \eta d$ with the dynamic viscosity 
$\eta = \SI{e-3}{\pascal\second}$, the presence of the FGF surface leads to effectively larger values of $\zeta$. Second, for our system, smaller particles come closer to the FGF surface, and thus they are strongly attracted by its stray field, which results in a reduction of their thermal fluctuations. For example, for a $\SI{270}{\nano\meter}$ particles an effective enhancement of the friction coefficient due to the presence of the FGF surface is estimated to be $25-30\%$, as follows from the reduction of the diffusion coefficient (cf. the inset of Figure 2) relative to the bulk diffusivity.  

\begin{figure}[ht]
\begin{center}
\includegraphics[width=\columnwidth,keepaspectratio]{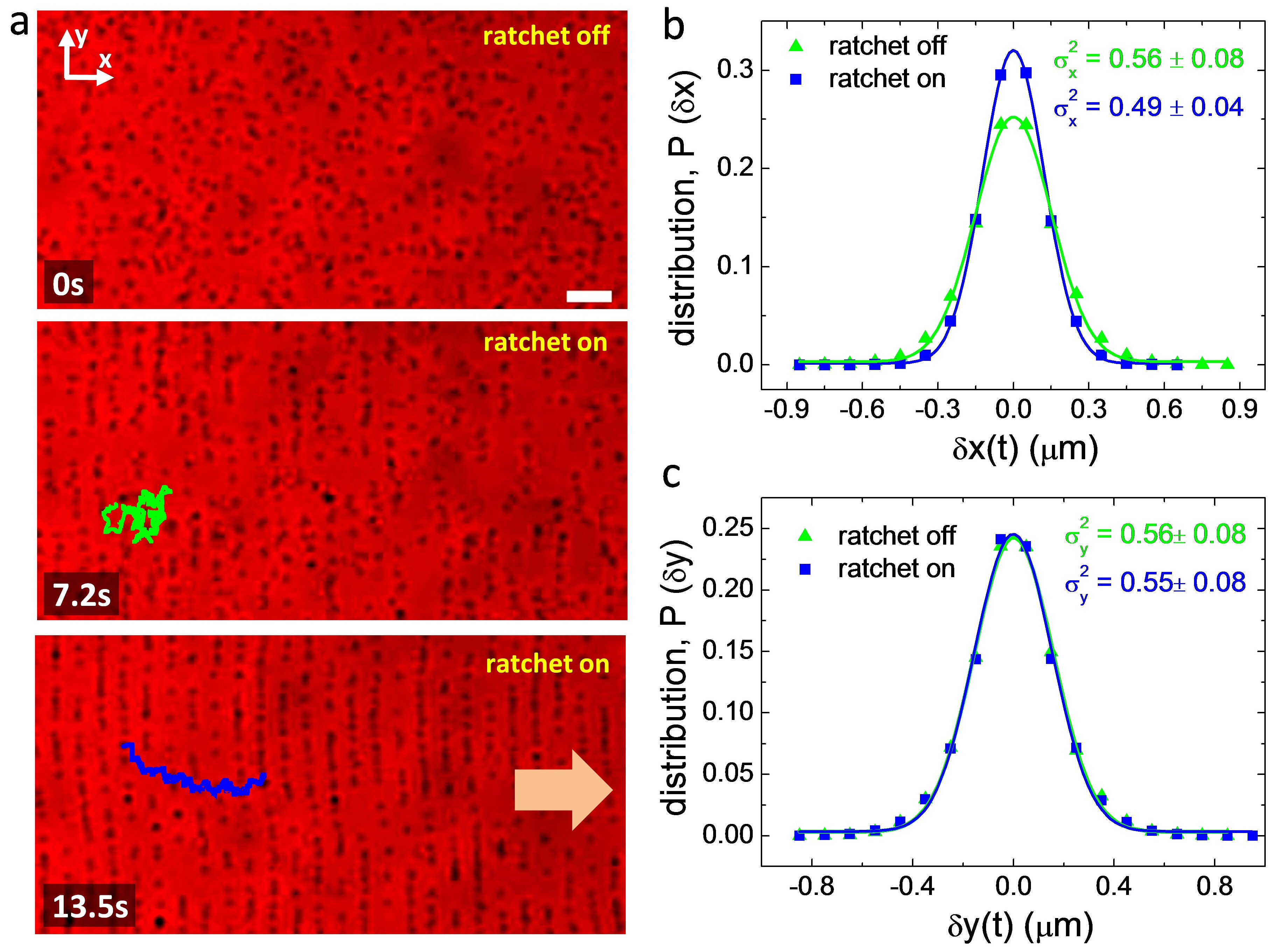}
\caption{(a) Sequence of snapshots showing the collective transport of $\SI{270}{\nano\meter}$ particles.
At time $t=\SI{7.2}{\second}$ the magnetic ratchet is switched on 
and all the nanoparticles are transported 
to the right (arrow at bottom) at an average speed $\langle v \rangle=\SI{2.6}{\micro\meter\per\second}$ (field parameters $f=\SI{1}{\hertz}$, $H_0=\SI{1200}{\ampere\per\meter}$, $\beta=-1/3$). The green (middle figure) and blue (bottom figure) lines are trajectories of a single particle.
Scale bar is $\SI{5}{\micro\meter}$ for all images, see also 
VideoS2 in the Supporting Information.
(b) and (c) Probability distributions of the
particle position perpendicular $P(\delta x)$ (b) and 
parallel  $P(\delta y)$ (c) to the BWs. 
The scattered points are experimental data 
while the solid lines are
Gaussian fits with $\sigma_x^2$ and $\sigma_y^2$ 
for the variance in the corresponding directions. 
In (b) the distribution was calculated by subtracting the 
drift as described in the text.}
\label{fig4}
\end{center}
\end{figure}

While we have analyzed the transport of single particles,
our magnetic ratchet enables also the collective motion for an 
ensemble of nanoparticles.
We demonstrate this feature in Figure \ref{fig4}(a), where a 
dense suspension of $\SI{270}{\nano\meter}$
particles are trapped and transported 
across the FGF surface,
see also corresponding VideoS2 in the Supporting Information. 
In the absence of applied field (ratchet off)
the colloidal suspension shows simple diffusive dynamics as illustrated by the 
Gaussian distribution of the displacements
perpendicular (along the $x$ axis) and parallel (along the $y$ axis)
to the magnetic stripes, Figures \ref{fig4}(b,c).
Here we used $P(\delta x)=(2\pi\sigma_x^2)^{-1}\exp(-\delta x^2/2\sigma_x^2)$ 
with the variance $\sigma_x^2(t)=\langle \delta x^2(t)\rangle$, $\delta x(t) = X(t) - \langle X(t)\rangle$ and $X(t)=x(t)-x(0)$ defined as earlier and considered at a sufficiently large time. The distribution $P(\delta y)$ is defined in a similar way, for which $\langle Y \rangle \equiv 0$, implying no mean drift and free diffusion in the $y$ direction.
We note that when the ratchet is off, $H_0=0$, the magnetic landscape $U_\text{m} \propto H_0$ becomes effectively flat and does not affect the dynamics of the particle. In this situation, the mean drift disappears, $\langle v \rangle =0$, and therefore $\langle X \rangle = \langle v \rangle t = 0$. As a result, when the ratchet is off, both distributions (greens lines in Figures \ref{fig4}(b,c)) are identical, with $\sigma_x^2=\sigma_y^2$ and thus featuring an isotropic free diffusion dynamics.

Upon application of the rotating
field, $H_0 >0$, the particles are immediately localized along the 
BWs, forming parallel chains and being consecutively transported 
across the magnetic platform.
The ratchet transport features a similar dispersion along the perpendicular direction as in the free case, as shown by the blue line in Figure \ref{fig4}(b).
Along the transport directions we observe 
a stronger confinement with a narrower distribution of displacement $P(\delta x)$, where the mean drift is subtracted by putting $\langle X(t)\rangle =\langle v \rangle t$.
VideoS2 in the supporting information shows the versatility of our magnetic ratchet approach, as now the nanoparticles can be easily 
trapped or released and transported to the right or left by just inverting 
the chirality of the rotating field, $H_x \to -H_x$, which keeps $H_0$ and $\beta$ unchanged. While these features have been previously reported for micro-scale systems~\cite{Gun05,Yel05,Ehr11,Don10}, our experimental realization proves its potential for further miniaturization 
of the transported elements and opens the doors towards applications in 
magnetic drug delivery systems using the higher surface to volume ratio of nanoparticles. 

To conclude, we have reported the controlled transport and 
diffusion enhancement of nanoparticles in 
a magnetic ratchet generated at the surface 
of a ferrite garnet film. 
In contrast to previous experimental works on microscopic particles 
confined in a rotating optical ring~\cite{Gri2006,Evs2008}, above patterned plasmonic~\cite{Sha2010} or lithographic~\cite{Mic18} landscapes,
our nanoparticles are trapped and controlled 
on an extended surface in absence of any topographic relief 
that can perturb the transport via steric interactions. 
The advantage of using the garnet film as a source of
magnetic background potential is that the domain wall motion
is essentially free from intermittent behavior and hysteresis. This
makes the cyclic displacements and the device performance fully
reversible. 
We have also reported in the past the giant diffusion in a magnetic ratchet of microscopic colloids~\cite{Tie2010}. However, the significant advantage of the present implementation is that the particle fluctuations can also be controlled along the direction of motion and remain completely independent of the deformation of the BWs in the lateral directions. This crucial difference has important implications for the design of channel-free nanofluidic devices where 
the colloidal motion can be confined to a straight line.
Further, our work goes beyond a previous one~\cite{Tie2016},
centered on trapping fluctuating nanoparticles along the BWs in an FGF. Here, we not only demonstrate 
the possibility to transport nanoscale objects 
at a well defined speed, but also that our magnetic ratchet system can be used to tune the diffusive properties of the particles, increasing their effective diffusion constant by almost an order of magnitude with respect to the previous case.
Another future avenue of this work is to investigate the dynamics of dense suspensions of interacting nanoparticles, and how collective effects alters the average particle flow. Increasing their density, however, requires a visualization procedure different from fluorescent labeling, to avoid artifacts during the tracking mechanism. While with relatively larger particles, the average speed of a colloidal monolayer was found to decrease with respect to the individual case~\cite{Tie12}, the presence of stronger thermal fluctuations for smaller particles may cause different non-trivial effects on the overall system dynamics.
Finally, our controlled enhanced diffusion in a transported ratchet may be used as a pilot system for fundamental studies related to
transport, diffusion and their complex relationship at the nanoscale.

\section{Materials and Methods}
\subsection{Experimental System and Methods}
We give further details on the sample preparation and 
experimental setup.
In order to
decrease the strong magnetic attraction,
we coat the FGF film
with a $h=\SI{1}{\micro\meter}$ thick
layer of a photoresist (AZ-1512 Microchem, Newton, MA),
i.e. a light curable polymer matrix,
using spin coating at \SI{3000}{rpm} for \SI{30}{\second}
(Spinner Ws-650Sz, Laurell).
Before the experiments, 
each type of particle is diluted in highly deionized water
and deposited above the
FGF, where they sediment
due to the magnetic attraction to the BWs.

External magnetic fields were applied via custom-made Helmholtz coils connected to two independent power amplifiers (AMP-1800, Akiyama), which are controlled by a wave generator (TGA1244, TTi). Particle positions and dynamics are recorded using an upright optical microscope (Eclipse Ni, Nikon) equipped with a 100 $\times$ 1.3 NA oil immersion objective and a CCD camera (Basler Scout scA640-74fc, Basler) working at $75$ frames per second. The resulting field of view is \num{65 x 48}~\si{\micro\meter^2}.

Videomicroscopy and particle tracking routines~\cite{Crocker1996} are used to extract the particle positions $\{x_i(t), y_i(t)\}$, with $i=1,\dots,N$,
from which the mean speed $\langle v \rangle$ is obtained performing both time and ensemble averages.
To calculate the mean square displacement and diffusion coefficient, we use $N$ trajectories  with length $l_\text{threshold}=200\,$frames, corresponding to a measurement time of $\Delta t=200/75=2.6\,$s. 

\subsection{Theoretical model}

The motion of paramagnetic colloidal particles above the FGF can be well interpreted within a simple model.
In an external magnetic field $\mathbf{H}$, a spherical magnetically polarizable particle of volume $V$ behaves as an induced magnetic dipole with the moment $\mathbf{m}=V \chi \mathbf{H}$ and the energy of interaction $U_\text{m}=-\mu_0 \mathbf{m}\cdot \mathbf{H}/2$ \cite{And02}, where $\chi$ is the effective susceptibility and $\mu_0=\SI{4\pi e-7}{\henry\per\meter}$ is the magnetic permeability of free space. The total magnetic field above the FGF $\mathbf{H}$ is given by the superposition $\mathbf{H}^\text{ac}+\mathbf{H}^\text{sub}$ of the external modulation of elliptic polarization, $\mathbf{H}^\text{ac}=H_0(\sqrt{1+\beta} \cos\omega t,0,-\sqrt{1-\beta} \sin\omega t)$, and the stray field of the garnet film $\mathbf{H}^\text{sub}\approx (4M_\text{s}/\pi)\exp(-kz)(\cos kx, 0, -\sin kx)$ \cite{Str13}, where $M_\text{s}$ is the saturation magnetization, $\omega =2 \pi f$ is the angular frequency, $k=2\pi/\lambda$ is the wave number, $H_0$ is the amplitude and $\beta$ is the ellipticity of the modulation.

Being interested in the transport properties across the magnetic stripes and evaluating the magnetic force exerted on the particle, $F(x,t)=-\partial_x U_m(x,t)$ with $U_\text{m}=-\mu_0 V \chi (\mathbf{H}^\text{ac}+\mathbf{H}^\text{sub})^2/2$, in the overdamped limit we obtain an equation of motion,
\begin{align}
\dot x(t) = \zeta^{-1}F(x,t) + \xi (t), \label{LE-gen}
\end{align}
with $F(x,t)= -\zeta\lambda f_\text{c}(\sqrt{1+\beta} \cos\omega t \cos k x - \sqrt{1-\beta} \sin\omega t \sin k x)$ and a random variable $\xi(t)$ taking account of thermal fluctuations. Here, $f_\text{c}=8 \mu_0 \chi V M_\text{s} H_0 \exp(-kz) / (\zeta\lambda^2)$, and $\xi(t)$ obeys the properties $\langle\xi(t)\rangle=0$, $\langle\xi(t)\xi(t')\rangle=2 D_0 \delta(t-t')$, where $D_0$ is the free diffusion coefficient, $k_\text{B}T$ is the thermal energy and $\zeta$ is the friction coefficient. By using $f_\text{c}$ and $D_0$ as fitting parameters, we numerically integrate Equation \ref{LE-gen} and evaluate the velocity of mean drift $\langle v \rangle$ and effective diffusion $D_\text{eff}$, which capture well the experimental data.

To gain further insights, we approximate the general model for an arbitrary $\beta$, Equation \ref{LE-gen}, by a more tractable special case  $\beta=0$, which corresponds to a traveling wave potential, $U_\text{m}(x,t)=U_\text{m}(x-v_0 t) \propto \cos(k(x-v_0 t))$. In the reference frame co-moving with the speed $v_0=\lambda f$, in terms of a new variable $u(t)=-x(t)+v_0 t$ we obtain, $\dot u(t)=\lambda f - \lambda f_\text{c} \sin ku +\xi(t)$. The deterministic velocity of mean drift is known to be \cite{Str13}:
\begin{align}
\langle v \rangle_{T=0} = \left\{\begin{array}{ll}
 \lambda f, & f \le f_\text{c}, \\
\lambda f - \lambda\sqrt{f^2-f_\text{c}^2}, & f > f_\text{c}.
\end{array} \right.
\label{vav-det}
\end{align}
Here, $f_\text{c}$ plays the role of the critical frequency that separates the low frequency domain with the maximum possible speed of mean drift,  $\langle v \rangle = v_0 \propto f$ ($f < f_\text{c}$), from the high frequency domain where its efficiency progressively drops down, $\langle v \rangle \propto f^{-1} < v_0$ ($f > f_\text{c}$).
For nanoparticles, thermal fluctuations are, however, inevitable, and we consider their thermal motion in the associated potential,
\begin{align}
\frac{V(u)}{k_\text{B}T}=-\frac{\lambda f}{D_0} u -\frac{\lambda f_c}{D_0 k}\cos ku, 
\label{potV}
\end{align}
which admits evaluation of the velocity of mean drift and effective diffusion \cite{Reim01,Bur09}
\begin{align}
\langle v \rangle & = \lambda f - \frac{D_0}{\lambda} \frac{1-\exp(-{\lambda^2 f}/{D_0})}{ \langle I_{\pm}(u)\rangle_u }, \label{vav} \\
D_\text{eff}  & = D_0 \frac{\langle I_{\pm}^2(u)I_{\mp}(u)\rangle_u}{\langle I_{\pm}(u)\rangle_u^{3}}.
\label{deff}
\end{align}
Here, ${I}_{\pm}(u)=\langle \exp [ \pm(V(u)-V(u\mp u'))/k_\text{B}T ] \rangle_{u'}$ are evaluated for potential (Equation \ref{potV}) and $\langle\cdot\rangle_u=\lambda^{-1}\int_0^{\lambda} \cdot \,\mathrm{d}u$ denotes the average over one wavelength of the landscape.\\

\textbf{\Large{Acknowledgment}}

R. L. S. acknowledges support
from the Swiss National Science Foundation grant
No. 172065. 
A. V. S. acknowledges partial support by Deutsche Forschungsgemeinschaft (DFG)
through grant SFB 1114, Project C01.
P. T. acknowledges support from the ERC
starting grant "DynaMO" (335040) and from MINECO
(FIS2016-78507-C2) and DURSI (FIS2016-78507-C2).

\begin{suppinfo}
Two experimental videos (.WMF)
showing the 
dynamics of magnetic colloidal particles
above the FGF film. 
\end{suppinfo}

\bibliography{bibliography}
\clearpage 
\end{document}